\documentclass[11pt,aps,nofootinbib,
floatfix,superscriptaddress]{revtex4}
\usepackage{graphicx}
\usepackage{epsf,amsmath,amsfonts,amssymb,amsbsy}
\usepackage[mathscr]{eucal}
\usepackage{hyperref}

\begin{document}

\title{Quantum origin of suppression for vacuum fluctuations of energy}

\begin{abstract}
{By example of a model with a spatially global scalar field, we show
that the energy density of zero-point modes is exponentially suppressed by an
average number of field quanta in a finite volume with respect to the energy
density in the stationary state of minimal energy. We describe cosmological
implications of mechanism. }
\end{abstract}

\author{Ja.V.Balitsky,}
\affiliation{Moscow Institute of Physics and Technology (State
University), Russia, 141701, Moscow Region, Dolgoprudny, Institutsky 9}

\author{V.V.Kiselev}
\email{Valery.Kiselev@ihep.ru}

\affiliation{Moscow Institute of Physics and Technology (State
University), Russia, 141701, Moscow Region, Dolgoprudny, Institutsky 9}
\affiliation{Russian State Research Center Institute for High Energy
Physics (National Research Centre Kurchatov Institute), Russia, 142281,
Moscow Region, Protvino, Nauki 1}


\maketitle

\section{Introduction and recapitulation of cosmological constant problem}

After primary speculations presented in \cite{Zeldovich:1968zz}, the
cosmological constant \cite{Weinberg:1988cp} is associated with a vacuum
energy density generated by zero-point modes of quantum fields
\cite{Andrianov:2007db,Akhmedov:2002ts}. In the framework of quantum field
theory, these quantum fluctuations are ordinary divergent and they should be
renormalized. In this respect, the relevant renormalization is related with
actual thresholds of energies, at which particles and forces contribute
significantly. Then, usually one supposes that some combinations of Planckian
scale and particle masses generate the energy density of vacuum
$\rho_\mathrm{vac}$, so that the maximal estimate corresponds to the greater
scale known in the real physics and it yields $\rho_\mathrm{vac}\sim
m_\mathrm{Pl}^4$, where the reduced Planck mass $m_\mathrm{Pl}\approx
2.4\cdot 10^{18}$ GeV is given by the Newton constant $G$ as $8\pi
G\,m_\mathrm{Pl}^2=1$. Such kind of estimates is in the direct conflict with
the value extracted from the cosmological data \cite{Ade:2013zuv} giving
$\rho_\mathrm{vac}\mapsto\rho_\Lambda=\Lambda^4$ at $\Lambda\sim 10^{-3}$ eV,
that is 30 orders of magnitude less than the Planck mass.

However, nobody can guarantee the the Planck scale defining the strength of
gravitational interaction has to establish a fundamental mass scale or energy
threshold relevant to the cosmological constant, of course. In this respect,
one could follow more realistic way by using the well accomplished
description of particle physics in the Standard Model. So, the direct
observation of Higgs boson allows us to evaluate the energy density of
electroweak vacuum from the effective potential of Higgs boson in terms of
masses of Higgs boson and W boson, that gives $\rho_\Lambda^\mathrm{EW}\sim
m_H^2m_W^2\sim (10^2\mbox{ GeV})^4$, which exceeds the contribution due to
the additional condensates in the Quantum ChromoDynamics by 8 orders of
magnitude, at least. Then, the magnitude of mismatching the scale of
cosmological constant would be significantly relaxed from 30 orders to 14
orders, that is still a big deal, no doubt. Notice, that in this approach to
the estimate of cosmological constant scale, one ignores an arbitrary
constant shift of Higgs potential. This shift can originate from physics of
other fields. In addition, at the observed mass value the Higgs potential can
lose its stability at very large fields below the Planckian range due to
effects of renormalization group, that could produce the tunnel decay of
present Universe to another Universe with a different vacuum.

Let us show that such the suppression can be explained due to a finite volume
effect for quantum fluctuations in an excited non-stationary state. For the
sake of clarity we exhibit the
mechanism by considering a time-dependent scalar field $\phi(t)$, which is
spatially global and free. The action in a finite physical volume $V_R$ is
given by expression
\begin{equation}\label{action}
    S=V_R\int dt\,\frac12\left(\dot \phi^2-m^2\phi^2\right),
\end{equation}
where $\dot\phi=d\phi/dt$ and $m$ is the field mass. The specific reference
frame of space-time  suggestively should be associated with the reference
frame of homogeneous component of cosmic microwave background radiation in
the Universe, so that time $t$ could correspond to the cosmic time of
Friedmann--Robertson--Walker--Lemaitre metrics. The meaning of $V_R$ is the
volume       wherein the fluctuations of field are causal, so that it can be
considered as spatially global, while an inhomogeneity is evaluated by
$|\nabla\phi|\sim\delta \phi/\lambda_c$, where $\lambda_c$ is the Compton
length, $\lambda_c=1/m$, and $\delta \phi$ denotes the field fluctuation. The
basic motivation and consideration are further considered in the field model
of (\ref{action}), so we ignore the influence of curved space-time on the main effect
for the moment. However, we return to the discussion of this issue in Section
\ref{III}.

Formally, action (\ref{action}) corresponds to the harmonic oscillator of
``frequency'' $m$ and ``inertial mass'' $V_R$. Then, the dimensionless operators
$$
    \hat Q=\frac{\phi}{\phi_0},\qquad \hat P=\frac{\dot \phi}{\dot\phi_0},
$$
at
$$
    \phi_0^2=\frac{1}{V_R m},\qquad \dot\phi_0^2=\frac{m}{V_R},
$$
define the operators of annihilation and creation for the spatially global
field quanta
$$
    \hat a=\frac{1}{\sqrt{2}}(\hat Q+i \hat P),\qquad
    \hat a^\dagger=\frac{1}{\sqrt{2}}(\hat Q-i \hat P),
$$
with the standard commutator
$$
    [\hat a,\hat a^\dagger]=1.
$$
The hamiltonian takes the form
\begin{equation}\label{H}
    \hat H=\frac12\, V_R (\dot\phi^2+m^2\phi^2)=
    \frac12\, m (\hat P^2+\hat Q^2)=m \left(\hat a^\dagger\hat a+
    \frac12\right).
\end{equation}

If the field is non-stationary excited, its quantum state can be considered as a
superposition of oscillatory coherent states, which minimize uncertainties in
the field $\phi$ and its rate $\dot\phi$. Let us consider the coherent state
$|\alpha\rangle$ with the average number of quanta $n$ in the volume $V_R$:
$$
    \hat a|\alpha\rangle=\alpha|\alpha\rangle,\qquad\alpha^*\alpha=n,\qquad
    \alpha=\frac{1}{\sqrt{2}}(Q_0+ i P_0).
$$
The averaged energy reads off
$$
    \langle E\rangle=\langle\alpha|\hat H|\alpha\rangle=m\left(n+\frac12\right).
$$
Usually, the minimal energy shift of this state from the minimum of potential
is referred to the energy level of zero-point mode,
\begin{equation}\label{Emin}
    \delta_\mathrm{min} E=\frac12\, m.
\end{equation}

So, it defines the quantity, which we call the bare cosmological constant,
\begin{equation}\label{bare}
    \rho_\Lambda^\mathrm{bare}=\frac{m}{2V_R}=
    \langle\mbox{vac}|\rho|\mbox{vac}\rangle=
    m^2\langle\mbox{vac}|\phi^2
    |\mbox{vac}\rangle=\langle\mbox{vac}|\dot\phi^2
    |\mbox{vac}\rangle.
\end{equation}
This shows that the finite volume sets non-zero fluctuations of spatially
global field.

If the fluctuations of field match to the Planckian scale,
$\langle\mbox{vac}|\phi^2|\mbox{vac}\rangle\sim m_\mathrm{Pl}^2$, then the
bare cosmological constant $\rho_\Lambda^\mathrm{bare}$ takes a huge value,
that constitutes the cosmological constant problem. On the other hand, if we
restrict ourselves by the range of Standard Model, then we would expect that
the mass of scalar field is given by the mass of Higgs boson $M$, while the
fluctuations of the field square are of the order of natural scale in the
electroweak physics, i.e. $M^2$. So, one could arrive to the estimate that is
consistent with the expectations of Standard Model, but it still would get a
huge value.

However,  in the next section we show that if the number of quanta for
$\phi(t)$ is not equal to zero, $n\neq 0$, then only a fraction of energy
shift from the potential minimum refers to the zero-point mode and the
fractional part of energy does originate from the suppressed vacuum
fluctuations, hence, the suppressed fraction corresponds to the vacuum energy
density, indeed.

\section{Suppression mechanism in action}

Let us find the fraction of zero-point mode in the energy shift from the
minimum of potential for the coherent state exactly. So, decomposing the
state into the sum of vacuum $|\mbox{vac}\rangle$ and the state
$|\mbox{quanta}\rangle$ with nonzero numbers of stationary field quanta
$$
    |\alpha\rangle=\mathcal A_\mathrm{vac}|\mbox{vac}\rangle+
    \mathcal A_q |\mbox{quanta}\rangle,
$$
we evaluate the average density of energy
\begin{equation}\label{average}
    \langle\alpha|\rho|\alpha\rangle=\left|\mathcal A_\mathrm{vac}\right|^2
    \langle\mbox{vac}|\rho|\mbox{vac}\rangle+
    \left|\mathcal A_q\right|^2
    \langle\mbox{quanta}|\rho|\mbox{quanta}\rangle,
\end{equation}
where the probability to find $k$ quanta in the coherent state is given by
the Poisson distribution,
$$
    \left|\mathcal A_k\right|^2=\frac{n^k}{k!}\,\mathrm{e}^{-n},
$$
while the free hamiltonian does not mix the stationary states with different
numbers of quanta, of course.

Therefore, the average density of energy, which is observed in the gravity,
is decomposed as
\begin{equation}\label{decompose}
    \langle\rho\rangle=\left|\mathcal A_\mathrm{vac}\right|^2\,
    \rho_\Lambda^\mathrm{bare}+\rho_q,
\end{equation}
at
\begin{equation}\label{vac}
    \left|\mathcal A_\mathrm{vac}\right|^2=\mathrm{e}^{-n},
\end{equation}
and $\rho_q$ being the energy density of nonzero-point modes that for the
coherent state equals
$$
    \rho_q=\frac{m}{V_R}\,\left(n+\frac12-\frac12\,\mathrm{e}^{-n}\right)=
    \rho_\Lambda^\mathrm{bare}\left(2n+1-\mathrm{e}^{-n}\right).
$$
Thus, the true energy density generated by zero-point modes in the coherent
state is suppressed and it is given by
\begin{equation}\label{true}
    \rho_\Lambda=\left|\mathcal A_\mathrm{vac}\right|^2\,
    \rho_\Lambda^\mathrm{bare}\mapsto \mathrm{e}^{-n}\,
    \rho_\Lambda^\mathrm{bare}.
\end{equation}
Relation (\ref{true}) remains valid generically by the order of magnitude not
only for the coherent state, but also for the most ordinary, non-exotic
states yielding $\left|\mathcal A_\mathrm{vac}\right|^2\sim \mathrm{e}^{-n}$.
Anyway, we can hold the relation for the probability of zero-point mode in
the quantum state as the definition of effective number of quanta in this
state. Moreover, if the fluctuations of field quanta are \textit{statistically
occasional}, then the probability with respect to the number of quanta has to
fit the \textit{Poisson distribution}, and  hence, the quantum state should be the
coherent state.

Let us stress that the described effect of vacuum fluctuations suppression in
the excited non-stationary state necessary involves the finite volume, but it
has no connection to the well know Casimir effect, which is also related with
a restricted volume. Indeed, the Casimir effect takes place due to the
relevant modification of zero-point modes in the state of \textit{minimal}
energy for the system of restricted volume, while setting the state of
minimal energy in our consideration will mean $n\to 0$ and the suppression
factor will disappear, that results in the ordinary situation for the Casimir
effect. In other words, the suppression factor becomes essential if the
system is excited to the non-stationary state, when the Casimir effect is
irrelevant, and vise versa, the Casimir effect takes place, when the
suppression under study is not in action. In this respect, the problem of
cosmological constant comes back in force, when we deal with the stationary
state of minimal energy, while near this state the Casimir effect represents
the evidence for the reality of vacuum energy.

The meaning of bare cosmological constant is the following: if the system
would be in the very vacuum state, the bare cosmological constant would be
the energy density in the system. i.e. in the empty vacuum without any
fields, particles and quanta, just vacuum fluctuations only. If the system is
not empty and it is occasionally excited, the actual vacuum fluctuations are
suppressed, that mean the suppression of observed cosmological constant in
such the word. In this way, we assume in our model that there is no any
different contribution to the cosmological constant, say, like some induced
terms of various nature, since those additional terms
cannot be suppressed in the same manner.

The decomposition of (\ref{decompose})--(\ref{true}) would remain formal, if
the field is held free and it does not interact, and then
$\rho_\Lambda^\mathrm{bare}$ would set the minimal density of energy, of
course, while we would observe the total density of field energy without a
possibility to extract the energy density of suppressed zero-point
fluctuations. However, if the field interacts, then it goes a non-trivial
evolution: the quanta can mix and transform into quanta of matter fields,
while the vacuum transfers itself into itself, i.e. into the \textit{vacuum},
since it is \textit{stable}, hence, the contribution of zero-point modes into
the energy density remains constant with the evolution, and this contribution
is much less than the bare term because of excited state of field, that makes
decomposition in (\ref{decompose})--(\ref{true}) to be observable for the
interacting field\footnote{The energy-momentum tensor with interactions acts
as the source of matter production, hence, it can cause the creation of
quanta from the vacuum. However, this effect cannot influence on the
contribution of zero-point modes themselves into its energy density,
surely.}. The suppressed contribution of zero-point modes is observed as the
cosmological constant in the presence of matter quanta.

In other words, the decomposition of (\ref{decompose})--(\ref{true}) is based
on quantum mechanics, and it is detectable. The detection suggests the
interaction of field system. Roughly speaking, the field quanta decay to
visible particles of matter, while the suppressed vacuum fluctuations form
the observable cosmological constant.

Let us look at the pressure of zero-point modes in order to justify their
vacuum status. So, the bare zero-point modes have got the energy
\mbox{$E^\mathrm{bare}=\frac12 m$} independent on the reference volume. It
means that the pressure is given by \mbox{$p^\mathrm{bare}=\partial
E^\mathrm{bare}/\partial V_R\equiv 0$,} that can be also calculated by means
of averaging the spatial components of energy-momentum tensor,
$$
    \langle\mbox{vac}|T_\alpha^\beta|\mbox{vac}\rangle=-\delta_\alpha^\beta\,
    p^\mathrm{bare}=-\delta_\alpha^\beta\,
    \frac12\, \langle\mbox{vac}|\left\{\dot\phi^2(t)-m^2\phi(t)\right\}
    |\mbox{vac}\rangle=0.
$$
In contrast, the true energy of suppressed vacuum fluctuations in volume
$V_R$
$$
    E_\mathrm{vac}=V_R\rho_\Lambda=\frac12\, m \,
    \left|\mathcal A_\mathrm{vac}\right|^2
$$
can get the correct dependence on the volume, if we put the vacuum density of
energy to be constant, that implies $\left|\mathcal A_\mathrm{vac}\right|^2/
V_R=\mbox{const.}$ and the pressure gets the value as it should be in the
vacuum,
$$
    p=-\frac{\partial E_\mathrm{vac}}{\partial V_R}=-\rho_\Lambda,
$$
hence,
$$
    \frac{\partial n}{\partial \ln V_R}=-1.
$$
Therefore, the reference volume exponentially declines with the growth of
quantum number $n$, and there is a maximal number corresponding to a minimal
volume of Planckian length.

This derivation of actual value for the parameter of vacuum state is
elementary, but it could be absolutely impossible, if we would ignore the
variation of suppression factor in contrast to the case of stationary
ground state.

Let us evaluate the relative inhomogeneity of field with respect to the
energy density. So,
$$
    \frac{|\nabla\phi|^2}{m^2\langle\phi\rangle^2}\sim
    \frac{m^2(\delta\phi)^2}{m^2\langle\phi\rangle^2}\sim \frac{\delta E}{E}
    \sim\frac{1}{\sqrt{n}}\ll 1.
$$
Therefore, the inhomogeneity is negligible, if the field is non-stationary
excited to a large value of quanta.

\section{Cosmology and model estimates\label{III}}
As we have already emphasized we have ignored effects due to a curved
space-time, while we have derived our mechanism for the suppressed
cosmological constant despite it is relevant to the system with gravity,
indeed. In this respect, we assume that the relevant quantities can enter as
the \textit{initial conditions} for the further evolution of system by taking
into account the gravitational expansion. So, the energy density of vacuum
remains constant, while the quanta and its energy densities follow the
transformations in accordance with relative field equations taking into
account for the gravity, too.

Nevertheless, we have to mention that in the literature there are
computations of energy density for the zero-point modes (ZPM) in a curved
background, that take into account the dependence on the space-time curvature
(see, for instance, the textbook by Birrell and Davies
\cite{Birrell:1982ix}). So, modern investigations in
\cite{Maggiore:2010wr,Hollenstein:2011cz} argue for the fact that in the
curved space-time, for instance, in the de Sitter space-time being close to
the space-time of inflation in the early Universe the zero-point modes
themselves produce the energy density that quadratically evolves with the
Hubble rate. We stress that such the effect means that the energy density of
ZPM is not the cosmological constant at all, since the emergent equation of
state (EOS) deviates from the vacuum equation of state, when the ratio of
pressure to the energy density equals $-1$, and, hence, zero-point modes
generate the form of dark energy. So, one gets the opportunity to evaluate
the relevant parameter of EOS for such the dark energy. The effect found in
\cite{Maggiore:2010wr,Hollenstein:2011cz} essentially changes the energy
density of ZPM if the Hubble rate $H$ exceeds the scale of ZPM energy density
in the limit of $H\to 0$, i.e. in the flat space-time. In this respect we
expect that this influence of space-time evolution on the ZPM energy density
could be suppressed as the energy density divided by second degree of Planck
mass and second degree of huge energy scale in the bare cosmological
constant. In addition, the effect found in
\cite{Maggiore:2010wr,Hollenstein:2011cz} corresponds to the local fields
considered in the whole space-time, including Hubble and super-Hubble
distances, when, for instance, the dynamics of light scalar fields would be
essential \cite{Hollenstein:2011cz}. In contrast, in our model we deal with
the global field in the limited volume, which is much less than the Hubble
volume, when the approximation of field with the ordinary machinery of
particle representation is very close to the exact solutions at such the
distances deeply inside the Hubble horizon. So, we hope that the dynamical
aspects of ZPM energy density would not be crucial for the scheme offered in
the present paper.

In this context, there are similar and more general arguments in favor of
situation, when the vacuum energy cannot be the constant value and, hence, it
would never represent the cosmological constant, since the energy of vacuum
in the curved space-time evolves due to the renormalization group equations
with the Hubble rate as the evolution parameter
\cite{Shapiro:2000dz,Shapiro:2001rh,Sola:2007sv,Shapiro:2009dh,Sola:2013gha}.
Making use of conformal anomaly and other constructions in models, such the
investigations \cite{EspanaBonet:2003vk,Gomez-Valent:2014rxa} argue for the
dynamical vacuum energy, that at present can be tested by precision data in
cosmology. Again, these studies deal with the dark energy, but not the
cosmological constant, which is considered in our paper.

In our treatment of suppression factor, the calculations concerning for
zero-point modes in the curved space-time would change an exact value of bare
cosmological constant, however, the mechanism itself starts to work if the
system is essentially excited to the occasional non-stationary state, that
yields the suppression factor of bare cosmological constant even in the
presence of space-time curvature, of course. So, in our study we hold the
standard point of view on the vacuum energy equivalent to the cosmological
constant and do not involve the dynamical treatment of vacuum energy evolving
with the Universe expansion. In principle, we see that the offered mechanism
can be implemented in the studies with the dynamical vacuum energy, too,
simply by the insertion of suppression factor for the evolving value of
vacuum energy, that could be considered in further developments of the
presented way, elsewhere.

Further, another aspect of curved space-time is the particle creation, say,
during the cosmological evolution \cite{Birrell:1982ix}. Such the creation
changes the energy density of matter by an additional term depending on the
square of primary density of particles or Hubble rate in the fourth degree,
that is typical for the quantum effects in the curved space-time. In this
respect, the additional term is suppressed as the primary density of energy
to the Planck mass in the fourth degree. It means that such the contribution
is negligible if the energy density is significantly below the Planckian
density, that is assumed for the Universe beyond the region of quantum
gravity, of course. Moreover, such the gravitational creation of matter does
not influence the initial cosmological constant established at the start of
evolution. Thus, we expect that our model can be considered in the
cosmological aspects.

In the form of expression (\ref{true}) the described mechanism rigorously
sets the quantum suppression of bare cosmological constant. It is relevant to
the cosmology because, at first, the spatially global scalar field could be
associated with the spatially global part of inflaton
\cite{Guth:1980zm,Linde:1981mu,Albrecht:1982wi,Linde:1983gd}, second, a
finite volume of causal fluctuations corresponds to a primary volume of
Universe at the inflation start, wherein the inflaton field can be considered
as a spacially global\footnote{The volume of causal fluctuations is not
equivalent to the Hubble volume determined by the initial density of energy.
The Hubble volume could be greater than the finite volume of causal
fluctuations, of course. Therefore, the inflaton can get a valuable
inhomogeneity in the initial Hubble volume.}. In this way, we can make
estimations in a simple manner, say, by setting the primary fluctuations
of field as
$$
    \langle\mbox{vac}|\phi^2|\mbox{vac}\rangle\sim m_\mathrm{Pl}^2\quad
    \Rightarrow\quad \frac{1}{V_R}\sim m\, m_\mathrm{Pl}^2,
$$
hence, the average number of spatially global field quanta is evaluated by
$$
    n=\ln\frac{\rho_\Lambda^\mathrm{bare}}{\rho_\Lambda}\sim
    275-2\ln\frac{m_\mathrm{Pl}}{m}\gg 1,
$$
since the inflaton is quite heavy, i.e. $m\sim 10^{13}$ GeV, and $n\sim 250$.

Because the mechanism should be accepted as actual and justified, the true
question appears: What is the magic number of $n$? Our studies show that the
answer can be found, for instance, in the framework of model with the
inflaton non-minimally coupled to the gravity, i.e. due to the interaction
term of lagrangian in the form
$$
    \mathcal L_\mathrm{int}=\frac12\,\xi\,{\phi}\,{m_\mathrm{Pl}}\,R,
$$
where $R$ is the scalar curvature of metric in the Jordan frame
\cite{Starobinsky:1980te,Mukhanov:1981xt,Kallosh:2013tua,Bezrukov:2007ep,Giudice:2014toa}.
In the Einstein frame, the inflation scale is $\Lambda_\mathrm{inf}\sim
10^{16}$ GeV, while the transformed inflaton gets the mass of the order of
$\Lambda_\mathrm{inf}/\xi$. The parameters obey the relation for the strong
coupling\footnote{We are going to present the deeper consideration
elsewhere.}
$$
    n\sim\xi\sim \frac{m_\mathrm{Pl}}{\Lambda_\mathrm{inf}}.
$$
Note that $\xi\gg 1$ results in a strong suppression of amplitude in a
spectrum of relict gravitational waves, if the inflaton potential is exactly
quadratic, when it satisfies the form of cosmic attractor for parameters of
inflation \cite{Kallosh:2013tua,Giudice:2014toa}. A valuable amplitude of
relict gravitational waves would point to that the potential should involve
some non-quadratic terms breaking the attractor predictions at $\xi\gg 1$.
This amplitude of primary gravitational waves could be unambiguously
extracted from the detection of B-modes of cosmic microwave background
radiation, if the foreground polarization generated by the dust is suppressed
in a region of detection. At present, the enforced amplitude of B-mode of
cosmic microwave background radiation detected by BICEP2 \cite{Ade:2014xna}
corresponds to such the secondary foreground produced by the measured dust
distribution as the Planck collaboration has reported in \cite{Adam:2014bub}.

Thus, the issue on the enigmatic value of $n$ is transformed into a reasoning
to fix the hierarchy of $\Lambda_\mathrm{inf}\ll m_\mathrm{Pl}$, in fact. Why
does the inflation involve two energy scales? An answer would solve the
cosmological constant problem
\cite{Weinberg:2000yb,Rubakov:1999aq,Shapiro:2000dz,Klinkhamer:2007pe} by
product, to our opinion.

\section{Discussion and generalizations}

Let us argue for the relevance of spatially global scalar field to the
cosmological constant. In the framework of quantum field theory the sum of
divergent contributions of any existing fields into the vacuum energy density
should be treated as a new independent global dimensional quantity with zero
charges of vacuum. Therefore, we can consider this quantity being reducible
from the vacuum expectation of appropriate scalar field $\phi$ by introducing
the contribution to the lagrangian in the form of $\phi\Lambda_0^3$ that
reproduces the cosmological constant at some $\phi\mapsto \langle
\phi\rangle=\phi_0$. Without the gravity, the value of cosmological constant
is irrelevant to the physics and it can take any value that corresponds to
the global shift symmetry $\phi\mapsto \phi+\phi_c$, while the action can
contain any scalar terms dependent on $\partial_\mu\phi$ and trivial flat
potential of $\phi$. These properties are characteristic for the inflaton
field. This is gravity that is responsible for a generation of terms breaking
the global shift symmetry, particularly, a non-flat potential as well as the
kinetic term for $\phi$ that makes it to be the dynamical field of inflaton,
we believe.

Finally, we can straightforwardly generalize the mechanism to the calculation
of vacuum energy for the nonhomogeneous scalar field. In this case, the bare
expression for the average tensor of energy and momentum
$$
    \langle\mbox{vac}|T_\mu^\nu|\mbox{vac}\rangle=
    \langle\mbox{vac}|\left\{\partial_\mu\phi\, \partial^\nu\phi
    -\frac12\,\delta_\mu^\nu\, (\partial\phi)^2
    +\frac12\,\delta_\mu^\nu\, m^2\phi^2\right\}|\mbox{vac}\rangle
$$
can be written as the integral
$$
    \langle\mbox{vac}|T_\mu^\nu|\mbox{vac}\rangle=
    \int\frac{d^4k}{(2\pi)^4}\,\frac{i}{k^2-m^2+i0}\,
    \left\{k_\mu k^\nu-\frac12\,\delta_\mu^\nu\,(k^2-m^2)
    \right\}.
$$
After the Wick rotation $k_0=ik_4$ to the Euclidean space, wherein
$k^2=-k_E^2$, we get
$$
    \langle\mbox{vac}|T_\mu^\nu|\mbox{vac}\rangle=
    \int\frac{d^4k_E}{(2\pi)^4}\,\frac{1}{k_E^2+m^2}\,
    \left\{-k_{E\mu} k_E^\nu+\frac12\,\delta_\mu^\nu\,(k_E^2+m^2)
    \right\},
$$
and the isotropic integration makes the replacement
$$
    k_{E\mu} k_E^\nu\mapsto \frac14\,k_E^2\,\delta_\mu^\nu,
$$
resulting in the expected divergent expression with the vacuum signature of
$\delta_\mu^\nu$ in the tensor structure,
$$
    \langle\mbox{vac}|T_\mu^\nu|\mbox{vac}\rangle=\delta_\mu^\nu\,
    \frac14
    \int\frac{d^4k_E}{(2\pi)^4}\,\frac{1}{k_E^2+m^2}\,
    \left\{k_E^2+2m^2\right\}.
$$
At this stage we can use the introduction of effective number of quanta with Euclidean
four-momentum $k_E$, $n=n(k_E^2)$ in order to get the true expression for the
energy density of suppressed vacuum fluctuations,
\begin{equation}\label{vactrue}
        \rho_\Lambda=\frac14
    \int\frac{d^4k_E}{(2\pi)^4}\,\frac{\mathrm{e}^{-n(k_E^2)}}{k_E^2+m^2}\,
    \left\{k_E^2+2m^2\right\}.
\end{equation}
This value is finite, if $n$ increases with $k_E^2$, say, polynomially. So,
setting the ansatz of linear dependence
$$
    n(k_E^2)=\bar n+\frac{k_E^2}{\bar \Lambda^2}
$$
we find
$$
    \rho_\Lambda=\frac{1}{32\pi^2}\,\mathrm{e}^{-\bar n}\left\{
    \bar\Lambda^4+m^2\bar\Lambda^2-m^4\left(\ln\frac{\bar\Lambda^2}{m^2}-
    \gamma_E+\mathcal O\left(\frac{m^2}{\bar\Lambda^2}
    \ln\frac{\bar\Lambda^2}{m^2}\right)\right)\right\},
$$
where $\gamma_E\approx0.5772$ is the Euler gamma. In the limit of $\bar
n\mapsto n$ given in the case of spatially global field, the cosmological
constant gets the leading term of $\bar\Lambda^4$ by the virtual modes and
subleading contribution of $m^2\bar\Lambda^2$ analogous to the expression
derived for the zero-point modes of global
field\footnote{If $\bar \Lambda\mapsto \Lambda_\mathrm{inf}$, then one can
expect that $\bar \Lambda^4\sim m^2 m_\mathrm{Pl}^2$ yielding $m\sim
\Lambda_\mathrm{inf}^2/m_\mathrm{Pl}\sim 10^{13}$ GeV.} at
$\bar\Lambda\mapsto m_\mathrm{Pl}$. Anyway, the suppression factor gets the
form of
exponentiating the effective number of quanta for the spatially global field.
Thus, the quantum description of vacuum energy density in the non-stationary
state shows the justified difference from the naive expectations on the
cosmological constant formed by fluctuations of the zero-point modes, i.e.
the bare cosmological constant.

Note, that the energy density in the model of growing $n(k_E^2)$ formally
becomes infinite, unless we introduce an evident cut off:
$$
    V_R\int\limits_0^{\Lambda_\mathrm{cut}}\frac{d^3k}{(2\pi)^3}\,n(k^2)
    \sqrt{k^2+m^2}=E_\mathrm{tot},
$$
by setting a finite total energy $E_\mathrm{tot}$ in the reference volume.
Nevertheless, our consideration remains valid in the case of
$\Lambda_\mathrm{cut}\gg \bar \Lambda$.

Evidently, the offered mechanism looks to be not straightforwardly effective
in the case of any fermionic field, when the occupation number takes only two
values: zero and unit. However,  the \textit{global} fermionic field is not
relevant to the cosmological constant, of course. At this point, it is
important to note that we assign the bare cosmological constant to the sum
over \textit{all} contributions of physical fields in the word including
fermionic fields, and, say, the Higgs boson, quark-gluon condensates and so
on. This is the gravity who is the reason why \textit{the bare cosmological
constant is transformed into the real dynamical scalar field} with the vacuum
quantum numbers of charges, i.e. into the inflaton, which is non-stationary
excited from the state of minimal energy in the finite volume. Then, by means
of the quantum-mechanical craft the excitation produces the suppression of
bare cosmological constant.

We address this problem on various sources of total cosmological constant in
new paper \cite{Balitsky:2014csa}, wherein we argue on the pseudo-Goldstone
nature of inflaton with respect to the global shift of energy scale of vacuum
energy density. In this mechanism, the primary cosmological constant induced
by all of actual contributions is matched to the bare cosmological constant
of inflaton. This matching lifts the mentioned problem on the copious
ingredients of total cosmological constant, we hope.

This work is supported by Russian Foundation for Basic Research, grant \#
14-02-00096.


\bibliography{bibmech-fin}

\end{document}